\documentclass[letterpaper,aps,prb,twocolumn,showpacs,superscriptaddress]{revtex4}
\usepackage{graphicx}
\usepackage{feynmf}
\usepackage{amsmath}
\pdfoutput=1

\begin{document}

\title{Transverse spin diffusion in ferromagnets}

\author{Yaroslav Tserkovnyak} 
\affiliation{Department of Physics and Astronomy, University of California, Los Angeles, California 90095, USA}
\author{E. M. Hankiewicz}
\affiliation{Institut f{\"u}r Theoretische Physik und Astrophysik, Universit{\"a}t W{\"u}rzburg, 97074 W{\"u}rzburg, Germany}
\author{Giovanni Vignale}
\affiliation{Department of Physics and Astronomy, University of Missouri, Columbia, Missouri 65211, USA }

\begin{abstract}
We discuss the dissipative diffusion-type term of the form $\mathbf{m}\times\nabla^2\partial_t\mathbf{m}$ in the phenomenological Landau-Lifshitz equation of ferromagnetic precession, which describes enhanced Gilbert damping of finite-momentum spin waves. This term arises physically from itinerant-electron spin flows through a perturbed ferromagnetic configuration and can be understood to originate in the ferromagnetic spin pumping in the continuum limit. We develop a general phenomenology as well as provide microscopic theory for the Stoner and $s$-$d$ models of ferromagnetism, taking into account disorder and electron-electron scattering. The latter is manifested in our problem through the Coulomb drag between the spin bands. The spin diffusion is identified in terms of the transverse spin conductivity, in analogy with the Einstein relation in the kinetic theory.
\end{abstract}


%75.30.Ds	Spin waves (for spin-wave resonance, see 76.50.+g)
%72.25.-b	Spin polarized transport
%76.50.+g	Ferromagnetic, antiferromagnetic, and ferrimagnetic resonances; spin-wave resonance
%75.45.+j		Macroscopic quantum phenomena in magnetic systems

\date{\today}
\maketitle

\section{Introduction}

The problem of spin diffusion through conducting ferromagnetic medium attracted much attention over several decades.\cite{halperinPR69,leggettJPC70,singhPRB89sd,sobolevJAP94,qianPRL02,mineevPRB04} Semiclassical spin transport in the presence of a weak magnetic field $\mathbf{H}$ can be captured by the conventional diffusion equation (neglecting spin relaxation):
\begin{equation}
\partial_t\mathbf{S}=\mathbf{H}\times\mathbf{S}+D\nabla^2\mathbf{S}\,,
\label{sd0}
\end{equation}
where $D$ is the diffusion coefficient and $\mathbf{H}$ is the total effective field (omitting the gyromagnetic ratio), including the applied and exchange contributions. The first term on the right-hand side describes spin precession in the local field while the second term stands for the ordinary diffusion of spin density $\mathbf{S}$. Equation~(\ref{sd0}) is, however, not applicable to most realistic ferromagnets, whose spin interactions are characterized by a large exchange energy $\Delta_{\rm xc}$. In particular, when $\Delta_{\rm xc}$ is comparable to the Fermi energy (which is the case in transition metals), the spin precession in the exchange field cannot be treated in the diffusive transport framework. Furthermore, the time-dependent exchange field induces spin-pumping currents\cite{tserkovPRB03sv,tserkovRMP05} inside the ferromagnet with spatially inhomogeneous magnetization dynamics, which can considerably modify the self-consistent magnetic equation of motion. Here, we wish to elucidate the central role of such self-consistent dissipative spin currents, which govern the diffusion-like terms in the magnetic equation of motion in the limit of strong ferromagnetic exchange correlations.

This paper is a follow up to our previous work,\cite{hankiewiczPRB08} providing additional technical details and offering a broader phenomenological base. Apart from assuming strong exchange correlations limit, our phenomenological approach and the main results of the paper should not be sensitive to the microscopic details and do not rely on the specific model of the ferromagnetic material (such as the Stoner or an $s$-$d$ model, for example). The main goals of this paper are as follows: (i) to put  the results of Ref.~\onlinecite{hankiewiczPRB08} into a broader phenomenological perspective, (ii) to explicitly show that two quite different models|the spin-polarized itinerant electron liquid (treated in Ref.~\onlinecite{hankiewiczPRB08}) and the $s$-$d$ model|lead to the same phenomenology and can be treated in parallel, and (iii) to make direct contact with the spin-pumping theory.\cite{tserkovPRB03sv,tserkovRMP05}

To be specific, let us consider a continuous ferromagnetic medium, with the effective field and spin density initially pointing along the $\mathbf{z}$ axis. For weak excitations close to this state, we may try expanding the ensuing transverse spin-current density as\cite{leggettJPC70,mineevPRB04}
\begin{equation}
\mathbf{j}_i=-D^\prime\mathbf{z}\times\partial_i\mathbf{S}-D^{\prime\prime}\partial_i\mathbf{S}\,,
\label{sc0}
\end{equation}
which enters in the continuity equation:
\begin{equation}
\partial_t\mathbf{S}=\mathbf{H}\times\mathbf{S}-\sum_{i={x,y,z}}\partial_i\mathbf{j}_i\,.
\label{sd1}
\end{equation}
In the limit of vanishing ferromagnetic correlations, we recover Eq.~(\ref{sd0}) by setting $D^\prime\to0$ and $D^{\prime\prime}\to D$ in Eq.~(\ref{sc0}). Hereafter, we are focusing exclusively on the transverse spin dynamics and spin currents. The longitudinal spin flows are conventionally described in terms of the ordinary diffusion for spin-up and spin-down electrons with spin-dependent diffusion coefficients and spin-flip scattering between the up and down spin bands.\cite{valetPRB93} Understanding the transverse spin flows and dynamics requires more care, in part due to the inherently quantum-mechanical behavior in the case of a strong exchange field. When the magnetic excitation is driven by the self-consistent transverse field $\mathbf{h}=\mathbf{z}\times\mathbf{H}\times\mathbf{z}$, there should also be field-driven contributions to the transverse spin current (\ref{sc0}), such as $\mathbf{j}_i\propto\partial_i\mathbf{h}$.

The problem in fact simplifies in the limit of strong exchange correlations. We will in the following employ a mean-field view of ferromagnetism, where the collective spin dynamics are driven by the exchange field, $\mathbf{H}=-\Delta_{\rm xc}\mathbf{m}(\mathbf{r},t)$ (setting $\hbar=1$ throughout), parametrized by the local and instantaneous spin-density orientation, $\mathbf{m}=\mathbf{S}/S$, which has to be solved for self-consistently. Since we are only interested in the transverse spin dynamics, we set the magnitude of the spin density $S$ to be spatially and time independent. In the limit of large $\Delta_{\rm xc}$, the spin currents can be parametrized by $\mathbf{m}(\mathbf{r},t)$. We can thus proceed phenomenologically and expand $\mathbf{j}_i$ in spatial and time derivatives of $\mathbf{m}(\mathbf{r},t)$. For a static magnetic profile $\mathbf{m}(\mathbf{r})$, we have the familiar exchange spin flow
\begin{equation}
\mathbf{j}^\prime_i=-A\,\mathbf{m}\times\partial_i\mathbf{m}
\label{jp}
\end{equation}
(where $A$ is the material-dependent exchange-stiffness constant), which is the only first-order form allowed by spin-rotational and time-reversal symmetries. To avoid unnecessary complications, we will assume isotropic ferromagnet throughout this paper. Dynamics allow for dissipative spin-current contributions that break time-reversal symmetry:
\begin{equation}
\mathbf{j}^{\prime\prime}_i=-\eta\,\mathbf{m}\times\partial_i\partial_t\mathbf{m}\,.
\label{jpp}
\end{equation}
Focusing on linear deviations of $\mathbf{m}$ from the equilibrium, $\mathbf{m}^{(0)}=\mathbf{z}$, we omit terms such as $\partial_i\mathbf{m}\times\partial_t\mathbf{m}$.

According to the time-reversal property, the spin-current density (\ref{jp}) corresponds to the $D^\prime$ term in Eq.~(\ref{sc0}), while the spin-current density (\ref{jpp}) is analogous to the $D^{\prime\prime}$ term, although the latter two are certainly not identical. In fact, we wish to emphasize the striking difference between the diffusive picture for the spin currents, Eq.~(\ref{sc0}), on one side and Eqs.~(\ref{jp}) and (\ref{jpp}) on the other side, where we expand spin currents phenomenologically in terms of the time-dependent magnetic texture $\mathbf{m}(\mathbf{r},t)$. The latter approximation is specific to the limit of strong exchange correlations, where the nondissipative spin current, Eq.~(\ref{jp}), is determined by the instantaneous magnetic profile, while the dissipative spin current, Eq.~(\ref{jpp}), can be interpreted as quasiparticle spin pumping by the collective magnetic dynamics,\cite{tserkovRMP05} rather than ordinary spin diffusion. It is also instructive to draw analogy between coefficients $A$ and $\eta$ in Eqs.~(\ref{jp}), (\ref{jpp}) and the shear modulus and shear viscosity, respectively, in elasticity theory.

In the next section, we develop further the phenomenological grounds for Eqs.~(\ref{jp}) and (\ref{jpp}), before proceeding with microscopic calculations for the dissipative coefficient $\eta$ in Secs.~\ref{mc} and \ref{mf}. In Sec.~\ref{sp}, we discuss a spin-pumping interpretation of dissipative spin current (\ref{jpp}), before summarizing our work in Sec.~\ref{dc}.

\section{Phenomenology}

\subsection{Landau-Lifshitz theory}

The conventional starting point for studying ferromagnetic precession is the nondissipative Landau-Lifshitz (LL) equation\cite{landauBOOKv9}
\begin{equation}
\left.\partial_t\mathbf{m}\right|_{\rm LL}=\mathbf{H}^\ast\times\mathbf{m}\,,
\label{LL}
\end{equation}
where we define the effective field $\mathbf{H}^\ast$ as the functional derivative of the free energy:
\begin{equation}
\mathbf{H}^\ast\equiv\partial_{\mathbf{m}}F[\mathbf{m}]/S\,.
\label{H}
\end{equation}
In this Landau-Lifshitz phenomenology, which is applicable well below the Curie temperature, only the position-dependent direction of the magnetization is taken to be a dynamic variable, parametrizing the Free energy $F[\mathbf{m}(\mathbf{r})]$. The angular-momentum density $\mathbf{S}=S\mathbf{m}$ is assumed to be related to the magnetization by a constant conversion factor, the effective gyromagnetic ratio. (Abusing terminology, we say \textit{spin density} synonymously with \textit{angular-momentum density}.) Since in the common transition-metal ferromagnets the gyromagnetic ratio is negative, we wrote Eq.~(\ref{H}) with an extra minus sign in comparison to the standard definition, where $\mathbf{m}$ is taken to be the direction of the magnetization rather than the spin density. The right-hand side of Eq.~(\ref{LL}) is the phenomenological reactive torque on the spatially-resolved magnetic precession, which generalizes the simple Larmor precession of Eq.~(\ref{sd0}). Note that the dissipation power
\begin{equation}
P[\mathbf{m}(\mathbf{r},t)]\equiv-S\int d^3r\,\mathbf{H}^\ast\cdot\partial_t\mathbf{m}
\end{equation}
clearly vanishes according to Eq.~(\ref{LL}). We also easily verify that the time reversal (under which $t\to-t$, $\mathbf{m}\to-\mathbf{m}$, and $\mathbf{H}^\ast\to-\mathbf{H}^\ast$) leaves Eq.~(\ref{LL}) unchanged, as it should in the absence of dissipation. The only dissipative term we can write in the quasistationary limit (i.e., up to the first order in $\partial_t$), assuming spatially uniform and isotropic ferromagnet, is the so-called Gilbert damping:\cite{gilbertIEEEM04}
\begin{equation}
\left.\partial_t\mathbf{m}\right|_{\rm LLG}=\mathbf{H}^\ast\times\mathbf{m}-\alpha\,\mathbf{m}\times\partial_t\mathbf{m}\,,
\label{LLG}
\end{equation}
where $\alpha$ is a material-dependent dimensionless (Gilbert) constant. A typical experimental value for $\alpha$ turns out to be often of the order of $10^{-2}$ in various metallic ferromagnets, which means that it takes roughly $2\pi/\alpha\sim10$ precession cycles for an out-of-equilibrium magnetization to relax to a static equilibrium direction along $\mathbf{H}^\ast$. The Gilbert damping breaks time-reversal symmetry and causes a finite dissipation power:
\begin{equation}
P[\mathbf{m}(\mathbf{r},t)]=\alpha S\int d^3r\,(\partial_t\mathbf{m})^2\,.
\end{equation}
As a side comment, we note that an alternative, so-called Landau-Lifshitz damping term $\mathbf{m}\times\mathbf{H}^\ast\times\mathbf{m}$ is mathematically identical to the Gilbert damping $\mathbf{m}\times\partial_t\mathbf{m}$ in Eq.~(\ref{LLG}), up to an extra factor of $(1+\alpha^2)$ on the left-hand side of the equation.

The effective field $\mathbf{H}^\ast$ is in practice dominated by the applied magnetic field, magnetic crystal anisotropies, and magnetostatic (dipole-dipole) interactions. In the presence of spatial inhomogeneities, there is also exchange contribution to the free energy, which to the leading (quadratic) order in magnetic inhomogeneities can be written as\cite{landauBOOKv9}
\begin{equation}
F_{\rm xc}=\frac{A}{2}\int d^3r\left[(\partial_x\mathbf{m})^2+(\partial_y\mathbf{m})^2+(\partial_z\mathbf{m})^2\right]\,.
\label{Fxc}
\end{equation}
The corresponding effective field is
\begin{equation}
\mathbf{H}_{\rm xc}=-(A/S)\,\nabla^2\mathbf{m}\,,
\end{equation}
and the associated term in LL Eq.~(\ref{LL}) is
\begin{equation}
\left.\partial_t\mathbf{m}\right|_{\rm xc}=(A/S)\,\mathbf{m}\times\nabla^2\mathbf{m}\,.
\label{dmxc}
\end{equation}
This equation can also be formally written as
\begin{align}
\left.S\,\partial_t\mathbf{m}\right|_{\rm xc}&=-\sum_{i=x,y,z}\partial_i\mathbf{j}^\prime_i\,,\nonumber\\
\mathbf{j}^\prime_i&=-A\,\mathbf{m}\times\partial_i\mathbf{m}\,,
\label{ce}
\end{align}
which simply recovers our equilibrium spin current (\ref{jp}). We emphasize that this spin current does not depend on magnetic dynamics.

To summarize these preliminary considerations, the phenomenological LL equation describes collective magnetic precession driven by local effective fields as well as equilibrium spin currents. At this point, there is, however, a conspicuous asymmetry in the treatment of the dissipative correction to the LL equation, i.e., Gilbert damping (\ref{LLG}), which depends only on the local magnetic dynamics and thus does not involve spin currents. To overcome this ``discrepancy," we expand the dissipative terms to second order in spatial derivatives, generalizing Gilbert term to
\begin{equation}
\left.\partial_t\mathbf{m}\right|_{\rm diss}=-\alpha\,\mathbf{m}\times\partial_t\mathbf{m}+(\eta/S)\,\mathbf{m}\times\nabla^2\partial_t\mathbf{m}\,,
\label{gg}
\end{equation}
where $\eta$ is a new phenomenological parameter, characterizing spin-wave damping. Assuming spatial-inversion symmetry (under which $\partial_i\to-\partial_i$ and $\mathbf{m}\to\mathbf{m}$), prevents us from writing any phenomenological terms linear in spatial derivatives. Recall also that we are always assuming small perturbations with respect to a uniform equilibrium magnetization, so that all spatial and time derivatives must hit a single $\mathbf{m}$ (for example, a dissipative term of the form $\sum_i[\partial_i\mathbf{m}\cdot(\mathbf{m}\times\partial_t\mathbf{m})]\partial_i\mathbf{m}$ is disregarded since it is higher order in small deviations of $\mathbf{m}$). Additional quadratic terms would be allowed phenomenologically if, e.g., we developed our linearized theory with respect to an equilibrium magnetic texture, such as a domain wall or magnetic spiral. Some of such terms were discussed in Ref.~\onlinecite{forosPRB08}, which is beyond our present scope. Finally, we note that we wrote Eq.~(\ref{gg}) with no direct coupling to the effective field $\mathbf{H}^\ast$. We justify this by assuming that the ferromagnetic correlations are characterized by a very large energy scale $\Delta_{\rm xc}$, so that microscopic processes responsible for dissipation are driven by the collective variable $\mathbf{m}$, rather than directly by $\mathbf{H}^\ast$. In transition-metal ferromagnets, the internal exchange energy is of the order of eV, while the effective field $\mathbf{H}^\ast$ corresponds to microwave frequencies (i.e., at least three orders of magnitude smaller than the exchange energy). This means that when we excite magnetic dynamics by an external field, the microscopic degrees of freedom respond not to the small driving field but rather the much larger self-consistent exchange field parametrized by the time-dependent $\mathbf{m}$. For the same reason, the spin current in Eq.~(\ref{ce}) depends only on the magnetic profile $\mathbf{m}(\mathbf{r})$, irrespective of how it is created by applied fields.

The total dissipation power corresponding to Eq.~(\ref{gg}) now becomes
\begin{equation}
P[\mathbf{m}(\mathbf{r},t)]=\int d^3r\left[\alpha S\,(\partial_t\mathbf{m})^2+\eta\,(\partial_i\partial_t\mathbf{m})^2\right]\,.
\end{equation}
Similarly to Eq.~(\ref{ce}), we can also write the $\eta$ term in Eq.~(\ref{gg}) in the form of the divergence of the spin-current density
\begin{equation}
\mathbf{j}^{\prime\prime}_i=-\eta\,\partial_i(\mathbf{m}\times\partial_t\mathbf{m})\,,
\label{spc}
\end{equation}
reproducing Eq.~(\ref{jpp}). We thus identified two contributions to the spin-current density: usual exchange spin current (\ref{ce}) and dissipative spin current (\ref{spc}), which we will later interpret as the dynamically-driven spin pumping.\cite{tserkovPRB03sv,tserkovRMP05} Spin current (\ref{spc}) can thus damp down spin-wave excitations even in the absence of any spin-relaxation scattering.\footnote{It is natural to also wonder about a possible additional spin current of the form $\mathbf{j}_i\propto\partial_i\partial_t\mathbf{m}$, which does not break time-reversal symmetry. Such spin current leads to a wave-vector-dependent correction to the effective gyromagnetic ratio, which is very small in practice. It parallels the structure of the spin pumping in magnetic nanostructures, which consists of the dominant dissipative piece of the form $\mathbf{m}\times\partial_t\mathbf{m}$ and a smaller piece of the form $\partial_t\mathbf{m}$. The latter merely causes a slight rescaling of the gyromagnetic ratio. While the dissipative piece of the spin pumping has been unambiguously established in a number of experiments,\cite{tserkovRMP05} the small correction to the gyromagnetic ratio is yet to be observed.} The latter is, however, believed to be the culprit for a finite Gilbert damping $\alpha$,\cite{hankiewiczPRB07} which relaxes uniform magnetic precession by transferring its angular momentum to the atomic lattice.

In the presence of dissipative currents (\ref{spc}), the relative linewidth of the spin-wave resonance\cite{heinrichAP93} is proportional to $\alpha+(\eta/S)q^2$, for the wave vector $q$. In the absence of the Gilbert damping $\alpha$, thus, the spectral width of the spin-wave excitation would vanish in the long-wavelength limit.\cite{halperinPR69}

\subsection{Mermin ansatz for spin current}

We now wish to establish a microscopic procedure for evaluating the dissipative component of the spin current, Eq.~(\ref{spc}). Ref.~\onlinecite{hankiewiczPRB08} adapted Mermin ansatz\cite{merminPRB70} for this purpose, which we will reproduce below. Microscopically, the spin-current density $\mathbf{j}_i$ is carried by conducting electrons responding to the mean-field exchange interaction
\begin{equation}
\hat{H}_{\rm xc}=-\Delta_{\rm xc}\,\mathbf{m}(\mathbf{r},t)\cdot\hat{\boldsymbol{\sigma}}/2
\label{Hxc}
\end{equation}
in the self-consistent single-electron Hamiltonian (which could stem, e.g., either from the coupling to the localized $d$ electrons in the $s-d$ model or the itinerant electron Stoner/LDA exchange). $\hat{\boldsymbol{\sigma}}$ is the vector of Pauli matrices, which defines the electron spin operator.

Let us for the moment view exchange interaction (\ref{Hxc}) as an external parametric driving field, not concerning with a self-consistent determination of $\mathbf{m}(\mathbf{r},t)$. In particular, we may allow for an instantaneous deviation of the electron spin density $\mathbf{s}$ from the exchange-field direction $\mathbf{m}$. This will allow us for a trick to find the ensuing spin flows, which is what we are after. The spin-density continuity equation corresponding to Hamiltonian (\ref{Hxc}) is
\begin{equation}
\partial_t\mathbf{s}=\Delta_{\rm xc}\,\mathbf{z}\times(s\,\mathbf{m}-\mathbf{s})-\partial_i\mathbf{j}_i\,.
\label{Sc}
\end{equation}
The equilibrium orientation of $\mathbf{m}$ is taken to be along the $z$ axis and we assume small-angle excitations, which do not modulate the magnitude of the spin density, $s=|\mathbf{s}|$. $\mathbf{s}$ here is the spin density of the \textit{conducting} electrons, which in, e.g., the $s-d$ model has to be distinguished from the \textit{total} spin density $\mathbf{S}$ that enters Eq.~(\ref{sd1}).

We next use the Mermin ansatz to relate the spin-current density $\mathbf{j}_i$ to the spin density $\mathbf{s}$:
\begin{equation}
\mathbf{j}_i=\sigma_\perp\Delta_{\rm xc}\,\partial_i(\mathbf{m}-\mathbf{s}/s)\,,
\label{Ma}
\end{equation}
where $\sigma_\perp$ is the transverse spin conductivity, to be evaluated later by the Kubo formula. Eq.~(\ref{Ma}) is analogous to Ohm's law for electric current density, with the expression on the right-hand side reminiscent of the gradient of the electrochemical potential. The physical reasoning behind ansatz (\ref{Ma}) is simple: there should be no dissipative spin currents in the static configuration, which corresponds to $\mathbf{s}(\mathbf{r})=s\,\mathbf{m}(\mathbf{r})$. The advantage in writing the spin current in this form is that $\sigma_\perp$ will now have to be evaluated in the limit of $(\mathbf{q},\omega)\to0$. Combining Eqs.~(\ref{Sc}) and (\ref{Ma}) will then give us the spin current to the linear order in $\mathbf{q}$ and $\omega$: exactly what we need to relate it to Eq.~(\ref{spc}) and read out $\eta$. In fact, it is sufficient to find $\Delta_{\rm xc}(\mathbf{m}-\mathbf{s}/s)\approx-\mathbf{z}\times\partial_t\mathbf{m}$ from Eq.~(\ref{Sc}), which is valid to the linear order in $\omega$ and zeroth order in $\mathbf{q}$, before putting it into Eq.~(\ref{Ma}) to finally find
\begin{equation}
\mathbf{j}_i=-\sigma_\perp\,\partial_i(\mathbf{z}\times\partial_t\mathbf{m})\,.
\label{Mj}
\end{equation}
Comparing this with Eq.~(\ref{spc}), we immediately identify $\eta$ with the transverse spin conductivity:
\begin{equation}
\eta=\sigma_\perp\,.
\label{etas}
\end{equation}
Equation (\ref{etas}) can be interpreted as an analog of the Einstein relation for transverse spin diffusion in strong ferromagnets.

\subsection{Transverse spin conductivity}

As is the case with the charge conductivity, it is convenient to evaluate the transverse spin conductivity in the velocity gauge. Namely, we eliminate the spin ``potential," corresponding to small magnetization deviations $\delta\mathbf{m}=\mathbf{m}-\mathbf{z}$ in Eq.~(\ref{Hxc}), by the SU(2) gauge transformation
\begin{equation}
\hat{\psi}(\mathbf{r},t)\to e^{i\Delta_{\rm xc}\int_{-\infty}^tdt^\prime\,\delta\mathbf{m}(\mathbf{r},t^\prime)\cdot\hat{\boldsymbol{\sigma}}/2}\hat{\psi}^\prime(\mathbf{r},t)\,,
\label{Ut}
\end{equation}
at the expense of introducing the SU(2) vector potential
\begin{equation}
\hat{A}_i=-\Delta_{\rm xc}\int_{-\infty}^t dt^\prime\,\partial_i\mathbf{m}(\mathbf{r},t^\prime)\cdot\hat{\boldsymbol{\sigma}}/2\,,
\end{equation}
which enters the kinetic part of the single-particle Hamiltonian as
\begin{equation}
\hat{H}_k=\sum_i(p_i-\hat{A}_i)^2/2m^\ast\,,
\label{Hk}
\end{equation}
where $p_i=-i\partial_i$ and $m^\ast$ is the electron's effective mass (assuming exchange-split parabolic bands). It is easy to verify that the effective field driving the spin current in velocity gauge (\ref{Hk}), $\hat{E}_i=-\partial_t\hat{A}_i$, is the same as the fictitious field $\hat{E}_i=-\partial_i\hat{V}$ in original length gauge (\ref{Hxc}). One caveat is in order: Eqs.~(\ref{Ut})-(\ref{Hk}) are only valid for an Abelian exchange potential, which would be the case if only one vector component of $\delta\mathbf{m}(\mathbf{r},t)$ was modulated (e.g., $\delta m_x$ or $\delta m_y$) in space and time. Such scenario is sufficient for our purpose, in order to establish the transverse spin conductivity entering Eq.~(\ref{Ma}).

Fourier transforming the electric field $\hat{E}_i$ in time, $\int dt\,e^{i\omega t}$, the usual relationship is obtained: $\hat{E}_i(\omega)=i\omega\hat{A}_i(\omega)$. We now proceed to construct the semiclassical transport equation for the spin current driven by a spatially homogeneous fictitious field  $\mathbf{E}_i=\mbox{Tr}[\hat{E}_i\hat{\boldsymbol{\sigma}}]=\Delta_{\rm xc}\partial_i\mathbf{m}$, to deduce the long-wavelength conductivity defined by Ohm's law\footnote{We need to remark here that the above gauge transformation does not affect the transverse spin current in the linearized theory.}
\begin{equation}
\mathbf{j}_i=\sigma_\perp\mathbf{E}_i\,.
\label{js}
\end{equation}
The semiclassical spin-current response, in the presence of the exchange splitting $\Delta_{\rm xc}$, with disorder and electron-electron scattering is given by\cite{damicoPRB00}
\begin{equation}
\partial_t\mathbf{j}_i+\Delta_{\rm xc}\,\mathbf{z}\times\mathbf{j}_i=\frac{n\mathbf{E}_i}{4m^\ast}-\mathbf{j}_i\left(\frac{1}{\tau_\perp^{\rm dis}}+\frac{1}{\tau_\perp^{ee}}\right)\,,
\label{te}
\end{equation}
where $n$ is the total equilibrium (conducting) electron density. The second term on the right-hand side of Eq.~(\ref{te}) describes spin-current relaxation, due to disorder and electron-electron scattering. Note that even in Galilean-invariant systems, spin-independent Coulomb interaction between electrons causes relaxation of a homogeneous spin current, in contrast to the ordinary current. Solving Eq.~(\ref{te}) at low frequencies, we recover Eq.~(\ref{js}) for the current component along $\mathbf{E}_i$, with\cite{singhPRB89sd,hankiewiczPRB08}
\begin{equation}
\sigma_\perp=\frac{n}{4m^\ast}\frac{\tau_\perp}{1+(\tau_\perp\Delta_{\rm xc})^2}\,,
\label{spg}
\end{equation}
where the total transverse spin scattering rate is defined by
\begin{equation}
\frac{1}{\tau_\perp}=\frac{1}{\tau_\perp^{\rm dis}}+\frac{1}{\tau_\perp^{ee}}\,.
\label{tauperp}
\end{equation}
In particular, in the limit of weak spin polarization and no electron-electron interactions, $\tau_\perp$ should reduce to the ordinary momentum scattering time $\tau$, and $\sigma_\perp$ to the quarter of the Drude conductivity $n\tau/m^\ast$.

\section{Microscopic calculation}
\label{mc}

\subsection{Spin-current autocorrelator}

In order to substantiate the preceding phenomenology, we need to establish the microscopic expressions for the involved scattering times, $\tau_\perp^{\rm dis}$ and $\tau_\perp^{ee}$. In the velocity gauge discussed in the previous section, the transverse spin conductivity is given, according to the Kubo formula, by the spin-current autocorrelation function:\cite{hankiewiczPRB08}
\begin{equation}
\sigma_\perp=-\frac{1}{4m^{\ast2} V}\lim_{\omega\to0}\frac{\Im m\langle\langle\sum_l\hat{\sigma}_{xl}p_{xl};\sum_l\hat{\sigma}_{xl}p_{xl}\rangle\rangle_\omega}{\omega}\,,
\label{Kf}
\end{equation}
where the summation is over all electrons in volume $V$ and
\begin{equation}
\langle\langle\hat{A};\hat{B}\rangle\rangle_\omega=-i\int_{0}^\infty dt\,e^{i(\omega+i0^+)t}\langle[\hat{A}(t),\hat{B}(0)]\rangle
\end{equation}
represents the Fourier-transformed retarded (Kubo) linear-response function for the expectation value of the observable $\hat{A}$ under the action of a classical field that couples linearly to the observable $\hat{B}$. $\Im m$ in Eq.~(\ref{Kf}) is inserted out of convenience, since the linear in $\omega$ response function is guaranteed to be imaginary. (The zeroth-order in $\omega$ correlator includes also the omitted ``diamagnetic piece" of the spin current in the velocity gauge.)

Assuming isotropic disorder (and for the moment no electron-electron interactions), the ladder vertex corrections to the conductivity vanish and we only need to evaluate the bubble diagram defined by the (single-particle) spin-dependent Green's functions
\begin{equation}
G^{R,A}_\sigma(\mathbf{p},\omega)=\frac{1}{\omega-p^2/2m^\ast+\mu_\sigma\pm i/2\tau_\sigma}\,,
\end{equation}
where $\sigma=\uparrow,\downarrow$ ($=\pm$) is the spin index along the $z$ axis, $\mu_\sigma=\mu+\sigma\Delta_{\rm xc}/2$ is the spin-$\sigma$ electron Fermi energy, $\mu$ is the chemical potential, and $\tau_\sigma$ is the spin-dependent disorder scattering time. In the Born approximation for dilute white-noise disorder, the scattering rate is proportional to the electron density of states, and we can write $\tau_\sigma=\tau\,\nu/\nu_\sigma$, where $\tau$ parametrizes the strength of the scattering potential, $\nu_\sigma$ is the spin-$\sigma$ band density of states, and $\nu=(\nu_\uparrow+\nu_\downarrow)/2$. A straightforward calculation then leads to\cite{hankiewiczPRB08}
\begin{equation}
\sigma_\perp=\frac{n}{4m^\ast}\frac{1}{\tau_\perp^{\rm dis}\Delta_{\rm xc}^2}\,,
\label{ss}
\end{equation}
in the strong exchange coupling limit, where
\begin{equation}
\frac{1}{\tau_\perp^{\rm dis}}=\frac{4}{3}\frac{\mu_\uparrow+\mu_\downarrow}{n\tau(\nu_\uparrow^{-1}+\nu_\downarrow^{-1})}
\label{tdis}
\end{equation}
identifies the disorder contribution to effective transverse spin scattering rate (\ref{tauperp}).

\subsection{Spin-force autocorrelator}

In the presence of electron-electron interactions, it is convenient to express the spin-\textit{current} autocorrelator (\ref{Kf}) in terms of the spin-\textit{force} autocorrelator. To this end, we use the equation of motion for the operators defining Kubo formula (\ref{Kf}) to find
\begin{equation}
\sigma_\perp=-\frac{1}{4m^{\ast2}\Delta^2_{\rm xc}V}\lim_{\omega\to0}\frac{\Im m\langle\langle\sum_l\hat{\sigma}_{xl}F_{xl};\sum_l\hat{\sigma}_{xl}F_{xl}\rangle\rangle_\omega}{\omega}\,,
\label{Kff}
\end{equation}
where $F_{xl}=\dot{p}_{xl}=-i[p_{xl},\hat{H}]$ is the force operator along the $x$ axis for the $l$th electron. Evaluated with respect to a uniform magnetization, $\mathbf{m}=\mathbf{z}$, the force operator $F_{xl}$ consists of two pieces: the disorder force and the electron-electron interaction force. Evaluating correlator (\ref{Kff}) in the clean limit to second order in Coulomb interactions, one finds for the transverse spin scattering rate:\cite{qianPRL02,hankiewiczPRB08}
\begin{equation}
\frac{1}{\tau_\perp^{ee}}=\Upsilon(p)\,m^\ast a_B^2r_s^4(k_BT)^2\,,
\label{tee}
\end{equation}
where $a_B$ is the Bohr radius, $T$  temperature, $k_B$ Boltzmann constant, $r_s$ the dimensionless Wigner-Seitz radius, and $\Upsilon(p)$ is a dimensionless function of the spin polarization $p=(n_\uparrow-n_\uparrow)/n$ ($n_s$ being spin-$s$ electron density), which was discussed in Refs.~\onlinecite{qianPRL02,hankiewiczPRB08}. Notice that scattering rate (\ref{tee}) has the Landau quasiparticle scaling with temperature. The finite-frequency modification of scattering rate (\ref{tee}) is, furthermore, accomplished by replacing $(2\pi k_BT)^2\to(2\pi k_BT)^2+\omega^2$.

\subsection{Spin-density autocorrelator}
\label{fe}

It is also possible to calculate the transverse spin diffusion directly, as a linear spin-density response to the transverse magnetic field. We will carry that out in Sec.~\ref{mf} for two popular mean-field models of ferromagnetism in metals: the Stoner and the $s-d$ models. In addition to offering an alternative approach to the problem, this derivation provides a justification for the preceding heuristic utilization of the Mermin ansatz.

Starting with the mean-field Hamiltonian for itinerant electrons
\begin{equation}
\hat{H}=\frac{p^2}{2m^\ast}+U(\mathbf{r})-\mu-\Delta_{\rm xc}\,\hat{\sigma}_z/2\,,
\label{Hf}
\end{equation}
and directly solving for the self-consistent spin-density response to a small driving magnetic field, we will derive in the next section the following general relation:
\begin{equation}
\eta=\frac{\Delta^2_{\rm xc}}{q^2}\lim_{\omega\to0}\frac{\Im m\,\tilde{\chi}_{+-}(\mathbf{q},\omega)}{\omega}\,,
\label{etasd}
\end{equation}
valid at long wavelengths, $q\to0$. The axially-symmetric (Kubo) spin-response function is defined by
\begin{equation}
\tilde{\chi}_{+-}(\mathbf{q},\omega)=-\frac{1}{2}\langle\langle s_+(\mathbf{r},t);s_-(\mathbf{r^\prime},0)\rangle\rangle_{\mathbf{q},\omega}\,,
\label{chimp}
\end{equation}
where $s_\pm=s_x\pm is_y$ is the transverse spin density of itinerant electrons. The disorder potential $U(\mathbf{r})$ entering Eq.~(\ref{Hf}) is, as before, taken to obey the Gaussian white-noise correlations:
\begin{equation}
\left\langle U(\mathbf{r})U(\mathbf{r^\prime})\right\rangle=\frac{1}{2\pi\nu\tau}\delta(\mathbf{r}-\mathbf{r^\prime})\,,
\label{U}
\end{equation}
where $\nu=(\nu_\uparrow+\nu_\downarrow)/2$ is the spin-averaged density of states at the Fermi level and $\tau$ is the characteristic scattering time.

\begin{widetext}
Writing the spin density $\mathbf{s}(\mathbf{r})=\mbox{Tr}\left[\boldsymbol{\hat{\sigma}}\hat{\rho}(\mathbf{r})\right]/2$ in terms of the electron density matrix $\rho_{\alpha\beta}(\mathbf{r})=\Psi^\dagger_\beta(\mathbf{r})\Psi_\alpha(\mathbf{r})$ in spin space, we proceed to evaluate $\tilde{\chi}_{+-}$ in the standard imaginary-time formalism. At temperature $T$, we have:
 \begin{equation}
\tilde{\chi}_{+-}(\mathbf{q},i\Omega_n)=-\frac{T}{2V}\sum_{\mathbf{p}\mathbf{p^\prime},m}\mathcal{G}_\downarrow(\mathbf{p}+\mathbf{q},\mathbf{p^\prime}+\mathbf{q};i\omega_m+i\Omega_n)\mathcal{G}_\uparrow(\mathbf{p^\prime},\mathbf{p};i\omega_m)\,,
\label{chiM}
\end{equation}
where
\begin{equation}
\mathcal{G}_\sigma(\mathbf{p},\mathbf{p^\prime};i\omega_m)=-\frac{1}{V}\int d^3\mathbf{r}d^3\mathbf{r^\prime}\int_0^{1/T} d\tau e^{-i\mathbf{p}\cdot\mathbf{r}+i\mathbf{p^\prime}\cdot\mathbf{r^\prime}+i\omega_m\tau}\left\langle\Psi_\sigma(\mathbf{r},\tau)\Psi^\dagger_\sigma(\mathbf{r^\prime},0)\right\rangle
\end{equation}
is the finite-temperature single-particle Matsubara Green's function. $\Omega_n=2n\pi T$ is the bosonic and $\omega_m=(2m+1)\pi T$ fermionic Matsubara frequencies, where $n$ and $m$ are integer indices.

The disorder-averaged Green's function is given by
\begin{equation}
\left\langle\mathcal{G}_\sigma(\mathbf{p},\mathbf{p^\prime};i\omega_m)\right\rangle=\frac{\delta_{\mathbf{p}\mathbf{p^\prime}}}{i\omega_m-\varepsilon_{\mathbf{p}\sigma}+i\,\text{sign}(\omega_m)/2\tau_\sigma}\,,
\label{Ga}
\end{equation}
where $\varepsilon_{\mathbf{p}\sigma}=p^2/2m^\ast-\mu-\sigma\Delta_{\rm xc}/2$. The analytic continuation of the Matsubara Green's functions into the retarded (advanced) Green's functions is accomplished by replacing $i\omega_m\to\omega\pm i0^+$ and ${\rm sign}(\omega_m)\to\pm$. According to our convention (\ref{U}), $\tau_\sigma=\tau\,\nu/\nu_\sigma$. Taking into account the vertex ladder corrections (as shown in Fig.~\ref{rpa}), we obtain for the disorder-averaged response function:
\begin{equation}
\tilde{\chi}_{+-}(\mathbf{q},i\Omega_n)=-\frac{T}{2V}\sum_m\frac{\sum_\mathbf{p}\mathcal{G}_\downarrow(\mathbf{p}+\mathbf{q};i\omega_m+i\Omega_n)\mathcal{G}_\uparrow(\mathbf{p};i\omega_m)}{1-(\xi/V)\sum_\mathbf{p}\mathcal{G}_\downarrow(\mathbf{p}+\mathbf{q};i\omega_m+i\Omega_n)\mathcal{G}_\uparrow(\mathbf{p};i\omega_m)}\,,
\label{chia}
\end{equation}
where $\xi=1/2\pi\nu\tau$ and by the Green's functions with a single wave-vector argument here we understand disorder-averaged propagators (\ref{Ga}). Inserting Eq.~(\ref{Ga}) into Eq.~(\ref{chia}) and performing an analytic continuation onto the real frequencies, it is straightforward to calculate $\tilde{\chi}_{+-}(\mathbf{q},\omega)$. Setting the temperature to zero and taking the $\omega\to0$ limit, we find:
\begin{equation}
\Im m\,\tilde{\chi}_{+-}(\mathbf{q},\omega)=\frac{\omega}{4\pi}\Re e\frac{\tilde{\chi}^{RA}-\tilde{\chi}^{AA}}{(1-\xi\,\tilde{\chi}^{RA})(1-\xi\,\tilde{\chi}^{AA})}\,,
\end{equation}
where $\tilde{\chi}^{XY}(\mathbf{q})=\int dp\,G^X_\downarrow(\mathbf{p})G^Y_\uparrow(\mathbf{p}-\mathbf{q})$ and $\int dp\equiv\int d^3\mathbf{p}/(2\pi)^3$ in three dimensions. All energies entering these Green's functions are set at the Fermi level. To the lowest order in $1/\tau\Delta$, we now obtain:
\begin{equation}
\Im m\,\tilde{\chi}_{+-}(\mathbf{q},\omega)=\frac{\omega}{8\pi}\left[\int dp\,A_\downarrow(\mathbf{p})A_\uparrow(\mathbf{p}-\mathbf{q})+4\xi\,\Im m\int dp\,G^A_\uparrow(\mathbf{p}-\mathbf{q})A_\downarrow(\mathbf{p})\int dp\,G^A_\uparrow(\mathbf{p}-\mathbf{q})\,\Re e\,G^R_\downarrow(\mathbf{p})\right]\,,
\label{chit}
\end{equation}
where $A_\sigma=-2\Im m\,G^R_\sigma$ is the spectral function.
\end{widetext}
The second term in Eq.~(\ref{chit}) is the vertex ladder correction, which is necessary for Eq.~(\ref{chit}) to give a meaningful result. In particular, the vertex correction cancels the spurious $q=0$ contribution of the first term, which would give $\alpha\sim1/\tau\Delta_{\rm xc}$. Finally, in the limit of $q\ll\pi\Delta_{\rm xc}/v_F$, we arrive at:
\begin{equation}
\Im m\,\tilde{\chi}_{+-}(\mathbf{q},\omega)=\frac{\mu_\uparrow+\mu_\downarrow}{3m^\ast\tau(\nu_\uparrow^{-1}+\nu_\downarrow^{-1})\Delta_{\rm xc}^4}\omega q^2\,.
\label{chiq}
\end{equation}
Using Eq.~(\ref{etasd}), this finally gives:
\begin{equation}
\eta=\frac{\mu_\uparrow+\mu_\downarrow}{3m^\ast\tau(\nu_\uparrow^{-1}+\nu_\downarrow^{-1})\Delta_{\rm xc}^2}\,,
\end{equation}
which agrees with Eqs.~(\ref{etas}), (\ref{spg}), and (\ref{tdis}) in the relevant here limit of $\tau_\perp^{-1}\ll\Delta_{\rm xc}$.

\section{Mean-field ferromagnetism}
\label{mf}

\subsection{Time-dependent LDA}
\label{lda}

In a spin-density-functional theory (s-DFT),\cite{capellePRL01,qianPRL02} the many-body problem of itinerant ferromagnetism is reduced to the single-electron Hamiltonian
\begin{align}
\hat{H}(t)=&\frac{p^2}{2m^\ast}+U(\mathbf{r})-\mu\nonumber\\
&-\left[\Delta_{\rm xc}\,\mathbf{m}(\mathbf{r},t)+\omega_0\mathbf{z}+\mathbf{h}(\mathbf{r},t)\right]\cdot\boldsymbol{\hat{\sigma}}/2\,.
\label{Hmf}
\end{align}
$\omega_0\ll\Delta_{\rm xc}$ is the ferromagnetic Larmor precession frequency in the presence of a uniform magnetic field applied along the $z$ axis. $\Delta_{\rm xc}\,\mathbf{m}(\mathbf{r},t)$ is the self-consistent exchange field, such that Hamiltonian (\ref{Hmf}) produces the correct spin-density response. Since we are ultimately interested in the equation of motion for the collective ferromagnetic dynamics, the spin-density response is all that is needed. In the local-density approximation (LDA) of the s-DFT, the exchange field follows the local and instantaneous magnetization direction $\mathbf{m}(\mathbf{r},t)$. $\mathbf{h}(\mathbf{r},t)$ is the external rf driving field, which we will treat perturbatively.

The time-dependent portion of the Hamiltonian is thus given by
\begin{equation}
\hat{H}^\prime(t)=-\int d^3\mathbf{r}\left[\Delta_{\rm xc}\,\delta\mathbf{m}(\mathbf{r},t)+\mathbf{h}(\mathbf{r},t)\right]\cdot\hat{\boldsymbol{\sigma}}/2\,.
\end{equation}
Since $\delta\mathbf{m}=\delta\mathbf{s}/S$ (where $\delta$ denotes small deviations from equilibrium), we have for the transverse spin component $s_+=s_x+is_y$:
\begin{equation}
s_+(\mathbf{q},\omega)=\tilde{\chi}_{+-}(\mathbf{q},\omega)\left[h_+(\mathbf{q},\omega)+\frac{\Delta_{\rm xc}}{S}s_+(\mathbf{q},\omega)\right]\,.
\end{equation}
The self-consistent response function to the rf field, $\chi_{+-}=s_+/h_+$, is thus
\begin{equation}
\chi^{-1}_{+-}(\mathbf{q},\omega)=\tilde{\chi}^{-1}_{+-}(\mathbf{q},\omega)-\frac{\Delta_{\rm xc}}{S}\,.
\end{equation}
In the LDA approximation, the problem thus trivially reduces to calculating the spin-spin response function for a noninteracting Hamiltonian with a fixed exchange field.

Let us in general write
\begin{equation}
\tilde{\chi}_{+-}(\mathbf{q},\omega)=\frac{S}{\left[\omega_r(\mathbf{q},\omega)+\Delta_{\rm xc}-\omega\right]-i\alpha(\mathbf{q},\omega)\omega}\,,
\label{chi}
\end{equation}
in terms of functions $\omega_r$ and $\alpha$ that are to be determined. The self-consistent response function then becomes:
\begin{equation}
\chi_{+-}(\mathbf{q},\omega)=\frac{S}{\left[\omega_r(\mathbf{q},\omega)-\omega\right]-i\alpha(\mathbf{q},\omega)\omega}\,.
\label{tchi}
\end{equation}
At $\mathbf{q}=0$, obviously $\omega_r(\omega)\equiv\omega_0$ and $\alpha(\omega)\equiv0$. This follows in general from the spin conservation in the presence of Coulomb interactions and arbitrary spin-independent potential $U(\mathbf{r})$. In this paper, we are most interested in the $q$-dependent damping function $\alpha(\mathbf{q},\omega)$, which can be identified by a microscopic evaluation of $\tilde{\chi}_{+-}(\mathbf{q},\omega)$. In the limit of strong exchange correlations, $\Delta_{\rm xc}\gg\omega_r$, we immediately obtain from Eq.~(\ref{chi}):
\begin{equation}
\alpha(\mathbf{q},\omega\to0)\approx\frac{\Delta_{\rm xc}^2}{S}\lim_{\omega\to0}\frac{\Im m\,\tilde{\chi}_{+-}(\mathbf{q},\omega)}{\omega}\,.
\label{a}
\end{equation}
In inversion-symmetric systems, the leading in $q$ spin-wave contribution to Gilbert damping is $\alpha(\mathbf{q},\omega\to0)=(\eta/S)q^2$, so that self-consistent response function (\ref{tchi}) corresponds to the dissipative term
\begin{equation}
\left.\partial_t\mathbf{m}\right|_{\rm diss}=(\eta/S)\mathbf{m}\times\nabla^2\partial_t\mathbf{m}
\label{eom}
\end{equation}
in Landau-Lifshitz Eq.~(\ref{LL}) of motion for the magnetic spin direction $\mathbf{m}(\mathbf{r},t)$. This is the desirable result and, according to Eq.~(\ref{a}), the microscopic expression for $\eta$ gives Eq.~(\ref{etasd}) of the previous section. In the next section, we will demonstrate that Eq.~(\ref{a}) is generic to mean-field treatment of conducting ferromagnets.

\subsection{$s-d$ model in RPA}
\label{sd}

It is also instructive to pursue a more basic description starting with a ferromagnetic lattice of localized $d$ electrons exchange-coupled to itinerant $s$ electrons. The corresponding Hamiltonian is
\begin{equation}
\hat{H}(t)=\hat{H}_0-\sum_i\left[J\boldsymbol{\mathcal{S}}_i\cdot\mathbf{s}(\mathbf{r}_i,t)+\boldsymbol{\mathcal{S}}_i\cdot\mathbf{h}(\mathbf{r}_i,t)\right]\,,
\end{equation}
where $\boldsymbol{\mathcal{S}}_i$ are local $d$ spins and $\hat{H}_0$ consists of the decoupled Hamiltonian for itinerant electrons, dc Zeeman Hamiltonian of the $d$ electrons, as well as the $d-d$ exchange and possible dipolar interactions. $\mathbf{h}$ is the applied rf field, which we take for simplicity to couple to the localized spin only. As long as the average exchange field experienced by the $s$ electrons is sufficiently strong and the magnetization is dominated by the $d$ electrons, we can disregard their direct rf coupling for our purpose. If also the Fermi wavelength is long in comparison to the $d$ lattice spacing, we will treat the electronic band structure in the effective-mass approximation, and also coarse grain the local spins: $\sum_i\boldsymbol{\mathcal{S}}_i\to\int d^3\mathbf{r}\,\boldsymbol{\mathcal{S}}(\mathbf{r})$ and $\boldsymbol{\mathcal{S}}_i\to(V/N)\boldsymbol{\mathcal{S}}(\mathbf{r})$, where $N/V$ is the density of $d$ sites.

\begin{figure*}[pth]
\includegraphics[width=\linewidth,clip=]{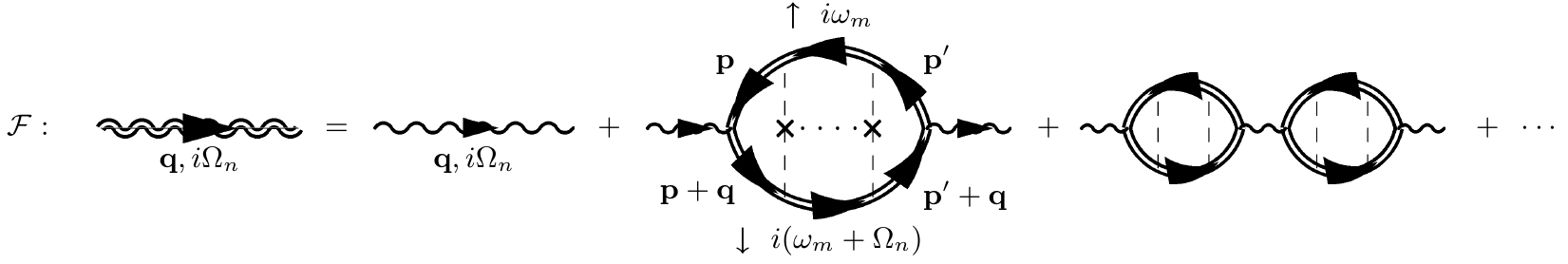}
\caption{In this RPA summation for the magnon propagator, Eq.~(\ref{mF}), the magnon self-energy is provided by the $s$-electron spin-response bubble, Eq.~(\ref{chiM}). Straight double lines denote disorder-averaged $s$-electron Green's functions and dashed lines describe vertex ladder corrections. Each bubble contains ladder corrections to the vertex function. Wavy lines are the $d$-electron spin propagators.}
\label{rpa}
\end{figure*}

Let us compute the spin-density response function for the $d$ lattice:
\begin{equation}
\chi_{+-}(\mathbf{r},\mathbf{r^\prime};t)=-\frac{1}{2}\langle\langle\mathcal{S}_+(\mathbf{r},t);\mathcal{S}_-(\mathbf{r^\prime},0)\rangle\rangle\,.
\label{Schi}
\end{equation}
For this purpose, it is convenient to define bosonic magnon operators:
\begin{equation}
a_\mathbf{p}=\frac{1}{\sqrt{2\mathcal{D}N}}\sum_ie^{-i\mathbf{p}\cdot\mathbf{r}_i}\mathcal{S}_{i+}\,,
\label{ad}
\end{equation}
which obey the canonical commutation relations, $[a_\mathbf{p},a^\dagger_\mathbf{p^\prime}]=\delta_{\mathbf{p}\mathbf{p^\prime}}$, close to the fully-magnetized ground state. To be specific, let us take the Heisenberg model for exchange coupling, so that in the ground state, $\mathcal{S}_{iz}=\mathcal{D}$, the $d$-orbital spin [assuming the applied dc magnetic field to point along the $-z$ direction, as in Eq.~(\ref{Hmf})]. The $d$-orbital Hamiltonian for magnon excitations close to the ground state can thus be written as:
\begin{equation}
\hat{H}_0=\sum_\mathbf{p}\varepsilon_\mathbf{p}a^\dagger_\mathbf{p}a_\mathbf{p}\,.
\label{H0}
\end{equation}
In terms of the magnon operators, we, furthermore, rewrite the $s-d$ exchange interaction as
\begin{align}
\hat{H}_{\rm xc}=&-\frac{J}{V}\sqrt{\frac{\mathcal{D}N}{2}}\sum_\mathbf{p}\left[a_{\mathbf{p}}^\dagger s_+(\mathbf{p})+a_\mathbf{p}s_-(-\mathbf{p})\right]\nonumber\\
&-J\sum_i\mathcal{S}_{iz}s_z(\mathbf{r}_i)\,,
\label{Hx}
\end{align}
where $s_\pm(\mathbf{p})=\int d^3\mathbf{r}\,e^{-i\mathbf{p}\cdot\mathbf{r}}s_\pm(\mathbf{r})$ is the Fourier-transformed transverse $s$-electron spin density. Approximating $\mathcal{S}_{iz}\approx\mathcal{D}$, we can combine the second term in Eq.~(\ref{Hx}) with $\hat{H}_0$ for the $s$ electrons, resulting in the longitudinal mean-field exchange field $\Delta_{\rm xc}=JS$, where $S=\mathcal{D}N/V$ is the averaged $d$-electron spin density. The spin-flop term of Eq.~(\ref{Hx}) can be expanded in terms of the electronic field operators $\Psi_\sigma$ as:
\begin{align}
\hat{H}_{\rm xc}^\prime=&-\frac{J}{V}\sqrt{\frac{\mathcal{D}N}{2}}\sum_{\mathbf{p}\mathbf{p^\prime}}\left[a^\dagger_\mathbf{p}\Psi^\dagger_\uparrow(\mathbf{p^\prime})\Psi_\downarrow(\mathbf{p^\prime}+\mathbf{p})\right.\nonumber\\
&\left.+a_\mathbf{p}\Psi^\dagger_\downarrow(\mathbf{p^\prime})\Psi_\uparrow(\mathbf{p^\prime}-\mathbf{p})\right]\,.
\label{Hp}
\end{align}

In the imaginary-time (Matsubara) formalism, response function (\ref{Schi}) can be written as:
\begin{equation}
\chi_{+-}(\mathbf{q},\tau)=-S\,\mathcal{F}(\mathbf{q},\tau)\,,
\label{chiF}
\end{equation}
in terms of the magnon Green's function
\begin{equation}
\mathcal{F}(\mathbf{q},\tau)=-\mathcal{T}\left\langle a_\mathbf{q}(\tau)a^\dagger_\mathbf{q}(0)\right\rangle\,,
\label{mF}
\end{equation}
using definition (\ref{ad}). Here, $\mathcal{T}$ is the time-ordering symbol. For noninteracting Hamiltonian (\ref{H0}),
\begin{equation}
\mathcal{F}_0(\mathbf{q},i\Omega_n)=\frac{1}{i\Omega_n-\varepsilon_\mathbf{q}}\,,
\end{equation}
where $\mathcal{F}(\mathbf{q},i\Omega_n)=\int_0^{1/T}d\tau e^{i\Omega_n\tau}\mathcal{F}(\mathbf{q},\tau)$ and $\Omega_n=2n\pi T$, as previously, is the bosonic Matsubara frequency. In order to calculate $\mathcal{F}$ in the presence of $s-d$ exchange (\ref{Hp}), we will sum up the bubble diagrams (which constitute an RPA approximation) shown in Fig.~\ref{rpa}. It is easy to recognize that each bubble (which is the magnon self-energy in this approximation) is just the itinerant electron spin-density response function that was already calculated in Sec.~\ref{fe}.

Summing up the diagrams in Fig.~\ref{rpa}, we thus obtain
\begin{equation}
\mathcal{F}^{-1}(\mathbf{q},i\Omega_n)=\mathcal{F}_0^{-1}(\mathbf{q},i\Omega_n)-\Sigma(\mathbf{q},i\Omega_n)\,,
\label{F}
\end{equation}
where the self-energy (the $s$-electron spin-response function)
\begin{equation}
\Sigma(\mathbf{q},i\Omega_n)=-J^2S\,\tilde{\chi}_{+-}(\mathbf{q},i\Omega_n)
\label{S}
\end{equation}
follows from Eq.~(\ref{chia}). Combining Eqs.~(\ref{chiF}), (\ref{F}), and (\ref{S}), we finally obtain (after analytic continuation onto real frequencies):
\begin{equation}
\chi^{-1}_{+-}(\mathbf{q},\omega)=\frac{\varepsilon_\mathbf{q}-\omega}{S}-J^2\tilde{\chi}_{+-}(\mathbf{q},\omega)\,.
\label{sdchi}
\end{equation}
Equation~(\ref{sdchi}) is actually quite trivial: it can be also obtained by treating the $d$-orbital magnetization dynamics and the associated $s-d$ torque in a mean-field approximation analogous to the preceding discussion of the Stoner model. Inverting Eq.~(\ref{sdchi}), we can write it in form (\ref{tchi}), after identifying
\begin{equation}
\omega_r(\mathbf{q},\omega)=\varepsilon_\mathbf{q}-SJ^2\Re e\,\tilde{\chi}_{+-}(\mathbf{q},\omega)
\end{equation}
and
\begin{equation}
\alpha(\mathbf{q},\omega)=SJ^2\frac{\Im m\,\tilde{\chi}_{+-}(\mathbf{q},\omega)}{\omega}\,,
\label{sda}
\end{equation}
which is, in fact, exactly the same as Eq.~(\ref{a}) in the low-frequency limit, using $\Delta_{\rm xc}=JS$. We should emphasize that, although $\chi_{+-}$ is defined in this section for a different physical system than $\chi_{+-}$ in Sec.~\ref{lda} (and thus, not surprisingly, is found to be somewhat different), the itinerant-electron response $\tilde{\chi}_{+-}$ is the same throughout the paper. The reason why the $q^2$ magnetic damping $\alpha(\mathbf{q},\omega)$ is identified in terms of the same quantity $\tilde{\chi}_{+-}$ in the two different models of ferromagnetism can be traced to our phenomenological identification of this damping in terms of the conducting-electron transverse conductivity, Eq.~(\ref{etas}). The latter is governed by the mean-field structure of the exchange field, irrespective of the microscopic origin of the ferromagnetic order.

Let us also note in the passing that, unlike the idealized Stoner model considered in the previous section, the $s-d$ magnetic damping may have a finite $q=0$ value even in the absence of any additional spin-dependent terms in the Hamiltonian. When the gyromagnetic ratios of the two electron species differ (ultimately stemming from some form of spin-orbit interaction), the total spin no longer precesses undamped in the uniform field, and the uniform transverse spin component can decohere in the presence of ordinary scalar disorder. Since Eq.~(\ref{sda}) corresponds to the magnetic field coupled to the $d$ electrons only, we implicitly set the $s$ electron $g$ factor to zero.

\section{Spin-pumping interpretation}
\label{sp}

It is illuminating to interpret the key result of this paper for the transverse spin diffusion of form (\ref{spc}) in terms of the spin pumping associated with a nonuniform magnetic dynamics in ferromagnetic bulk.\cite{tserkovRMP05} The ferromagnetic spin pumping was originally proposed in the context of magnetic multilayers with sharp normal-metal$\mid$ferromagnetic interfaces. This paper shows that analogous processes also take place in the continuous ferromagnetic medium.

\begin{figure}[pth]
\includegraphics[width=0.8\linewidth,clip=]{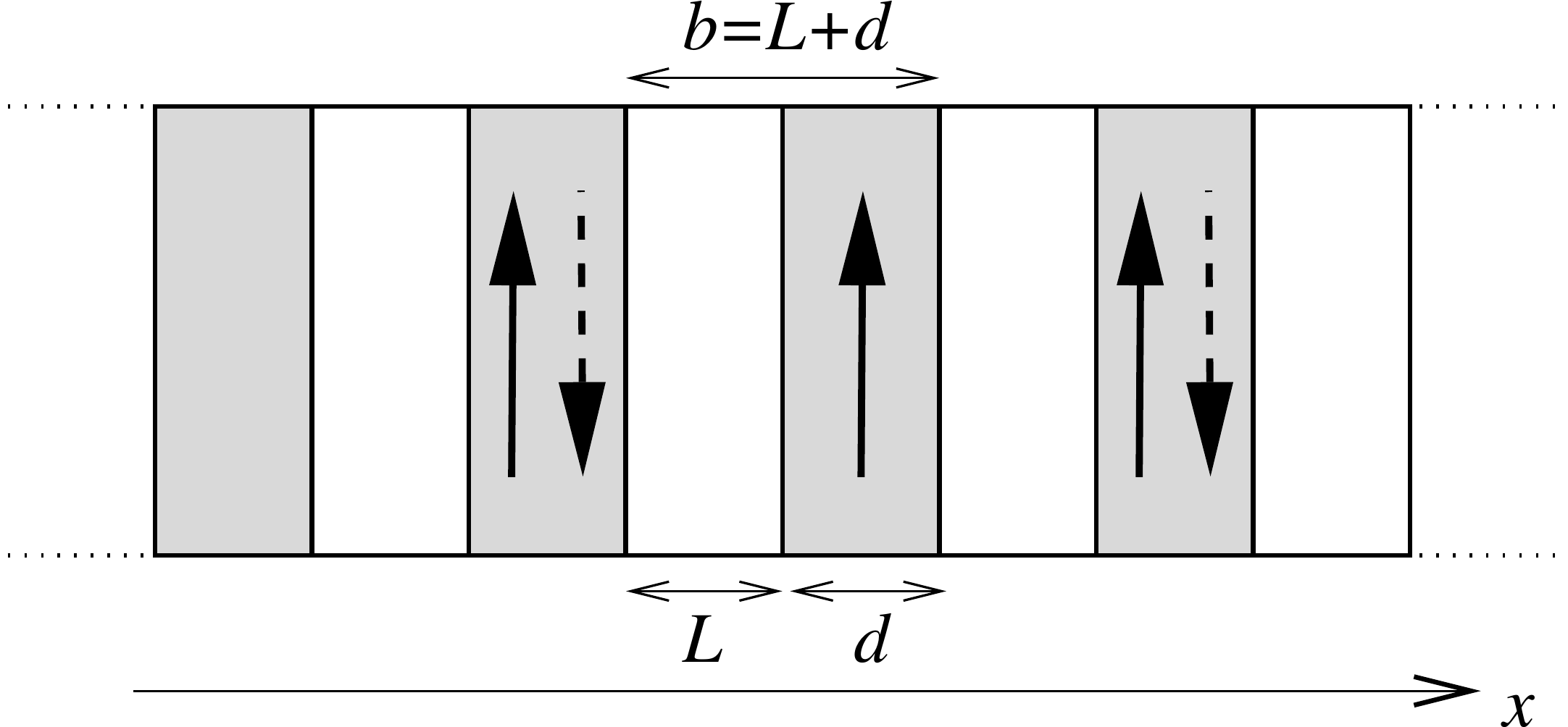}
\caption{A schematic view of the superlattice considered in the text: an F$\mid$N bilayer is repeated along the $x$ axis, with either ferromagnetic or antiferromagnetic alignment of the consecutive magnetic layers. The system is translationally invariant along the two remaining axes.}
\label{sl}
\end{figure}

To illustrate the direct connection between the transverse spin diffusion and the spin pumping, we consider a periodic stack of alternating F and N layers forming a two-component superlattice in the $x$ direction.\cite{tserkovRMP05} We treat the model depicted in Fig.~\ref{sl}, in which an F$\mid$N bilayer forms the unit cell with thickness $b=L+d$, where the normal-metal spacer of width $L$ separates the magnetic films of thickness
\begin{equation}
d\gg\lambda_{\rm sc}=v_F/\pi\Delta_{\rm xc}\,.
\label{lsc}
\end{equation}
The latter approximation allows us to neglect the transverse spin-current coherence between two interfaces of the same magnetic layer.\cite{tserkovRMP05} Translational invariance is assumed for simplicity in the lateral directions. We consider here collective spin-wave excitations, taking both the static and dynamic exchange couplings into account.\cite{tserkovPRB03sv}

The static (RKKY-like) exchange interaction between neighboring ferromagnetic layers is mediated by the dissipationless spin currents flowing through the normal-metal spacer.\cite{slonczewskiPRB89} We will parametrize the strength of this coupling by the corresponding precession frequency $\omega_{\rm xc}$ of a single ferromagnetic film that is exchange-coupled to a pinned film. In the presence of the magnetic dynamics, additional dissipative spin currents set in. Their origin lies in the spin pumping by the individual magnetic layers into the adjacent normal spacers, which at low frequencies is given by\cite{tserkovRMP05} $\mathbf{I}_s^{\rm pump}=(\hbar/4\pi)\tilde{g}_{{\rm N}\mid{\rm F}}^{\uparrow\downarrow}\mathbf{m}\times\partial_t\mathbf{m}$. $\tilde{g}_{{\rm N}\mid{\rm F}}^{\uparrow\downarrow}$ is the dimensionless spin-mixing conductance per unit area of the F$\mid$N interface (which is assumed to be real-valued, for simplicity). This interfacial spin pumping induces nonlocal spin transfer in magnetoelectronic circuits, which can in general be treated as a source term entering spin transport equations in normal and magnetic layers. In a collinear superlattice of Fig.~\ref{sl}, the problem simplifies considerably, because the spin-current vector $\mathbf{I}_s^{\rm pump}\propto\mathbf{m}\times\partial_t\mathbf{m}$ is transverse with respect to the magnetic alignment (in both ferromagnetic and antiferromagnetic cases), within the linear-response regime. This means that the spin current pumped by one ferromagnetic layer is either scattered back by the normal spacer and reabsorbed, or transmitted and absorbed by a neighboring layer, with no possibility to reach more distant neighbors, subject to condition (\ref{lsc}). Spin relaxation in normal spacers would only cause an overall increase in the effective Gilbert damping parameter of a uniform magnetic precession, and will thus be omitted, since our primary interest here is nonlocal damping effects. The problem of dynamic exchange between two adjacent ferromagnetic layers thus effectively reduces to the analogous effect in magnetic bilayers, which was studied in detail in Ref.~\onlinecite{tserkovPRB03sv}. In particular, the net spin pumping through a given normal spacer is $\propto\mathbf{m}_1\times\partial_t\mathbf{m}_1-\mathbf{m}_2\times\partial_t\mathbf{m}_2$, which reflects the dynamic spin injection in the opposite directions by the adjacent magnetic layers $\mathbf{m}_1$ and $\mathbf{m}_2$. Notice that the total pumping vanishes in the case of a perfectly synchronous precession, $\mathbf{m}_1(t)=\mathbf{m}_2(t)$.

Let us now put the static and dynamic exchange interactions into the equation of motion for small-angle spin dynamics of a multilayer with respect to an all-parallel configuration, $\mathbf{u}_{i}(t)=\mathbf{m}_{i}(t)-\mathbf{z}$. For long-wavelength excitations, it may be approximated as a continuous function $\mathbf{u}(x,t)$ of the coordinate $x$ normal to the interfaces. For the uniaxial effective field $H^\ast=-\omega_0\mathbf{z}$, the spin-wave dynamics obey the differential equation
\begin{align}
\partial_t\mathbf{u}=\left[\omega_0\mathbf{u}-\omega_{\rm xc}b^2\partial^2_x\mathbf{u}+\alpha\partial_t\mathbf{u}-\alpha^\prime b^2\partial^2_x\partial_t\mathbf{u}\right]\times\mathbf{z}\,,
\label{sle}
\end{align}
where we made the following definitions: $\omega_{\rm xc}=J_{\rm xc}/Sd$ and $\alpha^\prime=\Re e\,\mathcal{A}_{{\rm F}\mid{\rm N}\mid{\rm F}}^{\uparrow\downarrow}/4\pi Sd$. Here, $J_{\rm xc}$ is the exchange coupling between two consecutive magnetic layers and
\begin{equation}
1/\mathcal{A}_{{\rm F}\mid{\rm N}\mid{\rm F}}^{\uparrow\downarrow}=2/\tilde{g}_{{\rm N}\mid{\rm F}}^{\uparrow\downarrow}+e^2L\rho/\pi
\label{fnf}
\end{equation}
is the effective pumping resistance of the spacer. The first term on the right-hand side of Eq.~(\ref{fnf}) parametrizes the pumping strength of the individual interfaces, as was discussed above, and the second term is the ordinary ohmic resistance of the normal spacer (neglecting any spin relaxation), which backscatters the pumped currents and thus suppresses the dynamic exchange. The second spatial derivatives in Eq.~(\ref{sle}) reflect simply the difference of the static and dynamic exchange spin currents through two consecutive normal spacers (which themselves require a finite misalignment of the adjacent magnetic layers) in the continuum limit. The static Heisenberg coupling can be interpreted as the superlattice equivalent of the bulk exchange-stiffness parameter $A$ of Eq.~(\ref{Fxc}), which for the superlattice becomes $A=J_{\rm xc}b$. Both $\omega_{\rm xc}$ and $\alpha^\prime$ are sensitive to the normal-interlayer thickness $L$, vanishing in the limit $L\to\infty$. It follows from Eq.~(\ref{sle}) that the small-momentum, $q\ll b^{-1}$, spin-wave excitations of the superlattice, propagating perpendicular to the interfaces, $u\propto\exp[i(qx-\omega t)]$, obey the dispersion relation
\begin{equation}
\omega(q)=\frac{\omega_0+(bq)^2\omega_{\rm xc}}{1+i\left[\alpha+(bq)^2\alpha^\prime\right]}\,.
\label{oak}
\end{equation}
When $q\rightarrow0$, $\omega(q)$ reduces to the Larmor frequency $\omega_0$ of the individual magnetic layers because the static and dynamic exchange couplings vanish when the consecutive magnetic layers move coherently in phase. Equation~(\ref{oak}) holds up to momenta comparable to $b^{-1}$, when $bq$ has to be replaced by $2\sin(bq/2)$.

The matters are quite different for an antiferromagnetically-aligned superlattice, which is the ground state when, for example, $J_{\rm xc}<0$ and $H^\ast=0$. In this case, we have a more complex dispersion:
\begin{equation}
\omega(q)=-\omega_{\rm xc}\frac{\pm\sqrt{(bq)^2(1+\alpha^2)-4\alpha^2}+i[2\alpha+(bq)^2\alpha^\prime]}{1+\alpha^2+4\alpha\alpha^\prime+(bq)^2\alpha^{\prime2}}\,,
\label{obk}
\end{equation}
where plus and minus signs refer, respectively, to the modes with antisymmetric and symmetric dynamics in the adjacent layers for overdamped motion, and to the right- and left-propagating modes when the real part of $\omega(q)$ is significant. Note that now $\omega_{\rm xc}<0$, so that $\Im m\,\omega>0$, as required for a stable configuration. In the absence of bulk magnetization damping, $\alpha=0$, Eq.~(\ref{obk}) reduces to
\begin{equation}
\omega(q)=\frac{\pm(bq)\omega_{\rm xc}}{1\pm i(bq)\alpha^\prime}\,,
\label{obks}
\end{equation}
with linear dispersion and damping at small $q$. Equations~(\ref{obk}) and (\ref{obks}) can also be generalized to large momenta by replacing $bq$ with $2\sin(bq/2)$. Notice that in Eqs.~(\ref{sle}), (\ref{oak}), and (\ref{obks}), the dynamic coupling modifies the damping similarly to the way the static coupling affects the excitation frequency of the magnetic superlattice. Crystal and shape anisotropies on top of the simple effective fields assumed above might become important in real structures, and their inclusion is straightforward.

Let us now compare the damping $(bq)^2\alpha^\prime$ in Eq.~(\ref{oak}) with $\alpha(q)=(\sigma_\perp/S)q^2$ corresponding to Eq.~(\ref{ss}), which is the analogous quantity for the bulk. Keeping only the mixing conductance contribution to Eq.~(\ref{fnf}) and approximating\cite{tserkovRMP05} $\tilde{g}^{\uparrow\downarrow}\approx p^2_F/2\pi$ in terms of the characteristic Fermi momentum $p_F$ in the normal metal, we have for the $q$-dependent part of the damping:
\begin{equation}
\alpha(q)=(bq)^2\alpha^\prime\sim\frac{(b/\lambda_F)^2}{Sd}q^2\,,
\label{aqb}
\end{equation}
up to a numerical constant. At the same time, the bulk $\alpha(q)$, corresponding to Eq.~(\ref{ss}), can be written as
\begin{equation}
\alpha(q)\sim\frac{(\lambda_{\rm sc}/\lambda_F)^2}{Sl}q^2\,,
\label{aqa}
\end{equation}
which establishes a loose formal correspondence between the two results. Here,  $l=v_F\tau$ is the mean free path, $\lambda_F$ the Fermi wavelength, and the ferromagnetic coherence length $\lambda_{\rm sc}$ was defined in Eq.~(\ref{lsc}).

Comparing Eqs.~(\ref{aqb}) and (\ref{aqa}), we interpret the length scale $b\leftrightarrow\lambda_{\rm sc}$ to describe the longest distance over which ferromagnetic regions can communicate via spin transfer. The length scale $d\leftrightarrow\lambda$ characterizes momentum scattering relevant for spin transfer, which in the case of the superlattice with sharp interfaces corresponds to the magnetic film width $d$: Approximating $\tilde{g}^{\uparrow\downarrow}\approx p^2_F/2\pi$ above, we effectively took the normal spacers to be ballistic and, because of Eq.~(\ref{lsc}), the spin transfer does not penetrate deep into the ferromagnetic layers, making possible disorder scattering there irrelevant for our problem.

\section{Discussion and Outlook}
\label{dc}

Estimating the numerical value of the dimensionless $q^2$ damping, according to Eq.~(\ref{spg}),
\begin{equation}
\alpha(q)=\frac{\sigma_\perp q^2}{S}\sim\left(\frac{\mu_F/\Delta_{\rm xc}}{p_F/q}\right)^2\frac{\tau_\perp\Delta_{\rm xc}}{1+(\tau_\perp\Delta_{\rm xc})^2}\,,
\label{aq2}
\end{equation}
we can see that it will most likely be at most comparable or smaller than the typical $q=0$ Gilbert damping $\alpha\sim10^{-2}$, in metallic ferromagnets. Damping (\ref{aq2}) may, however, become dominant in weak ferromagnets, such as diluted magnetic semiconductors. We are not aware of systematic experimental investigations of the $q^2$ damping in metallic ferromagnets. $q^2$ scaling of relative linewidth was reported in Ref.~\onlinecite{fernandezJAP87} for the iron-rich amorphous Fe$_{90-x}$Ni$_x$Zr$_{10}$ alloys. However, we are not certain whether the strong damping observed there can be attributed to the mechanism discussed in our paper.

Another intriguing context where the physics discussed here can play out to be important is the current-driven nonlinear ferromagnetic dynamics in mesoscopic as well as bulk magnetic systems. The $q^2$ magnetic damping described by Eq.~(\ref{gg}) can be physically thought of the viscous-like spin transfer between magnetic regions precessing slightly out-of-phase. The obvious consequence of this is the enhanced damping of the inhomogeneous dynamics and thus the synchronization of collective magnetic precession. This phenomenon was predicted in Ref.~\onlinecite{tserkovPRB03sv} and unambiguously observed in Ref.~\onlinecite{heinrichPRL03}, in the case of the coupled dynamics of a magnetic bilayer: When the two layers are tuned to similar resonance conditions, only the symmetric mode corresponding to the synchronized dynamics produces a strong response, while the antisymmetric mode is strongly suppressed. It is thus natural to suggest that the $q^2$ viscous magnetic damping in the continuum limit may have far-reaching consequences for the current-driven nonlinear power spectrum as that measured in Ref.~\onlinecite{krivorotovPRB07}. This needs a further investigation.

The role of electron-electron interactions was manifested in our theory through the spin Coulomb drag, which enhances the effective transverse spin scattering rate (\ref{tauperp}). This becomes particularly important, in comparison to the disorder contribution to the transverse spin scattering, in the limit of weak magnetic polarization.\cite{hankiewiczPRB08}

We finally emphasize that the study in this paper was limited exclusively to weak linearized perturbations of the magnetic order with respect to a uniform equilibrium state. When the equilibrium or out-of-equilibrium magnetic state is macroscopically nonuniform, as is the case with, e.g., the magnetic spin spirals, domain walls, vortices, and other topological states, the longitudinal as well as transverse spin currents become relevant for the magnetic dynamics. The longitudinal spin currents lead to additional contributions to the spin-transfer torques, modifying the magnetic equation of motion. Such spin torques leading to the dissipative $q^2$ damping terms were discussed in Ref.~\onlinecite{forosPRB08}. These latter contributions to the magnetic damping are likely to dominate in strongly-textured magnetic systems.

\acknowledgments

We are grateful to Gerrit~E.~W. Bauer and Arne Brataas for stimulating discussions. This work was supported in part by the Alfred~P. Sloan Foundation (YT) and NSF Grant No. DMR-0705460 (GV).

\end{document}